\documentclass[aps,prl,reprint,showpacs,amsmath,amssymb]{revtex4-1}

\usepackage{graphicx}
\usepackage{microtype}

\renewcommand*{\vec}[1]{\boldsymbol{\mathbf{#1}}}
\DeclareMathOperator{\sgn}{sgn}
\newcommand*\diff{\mathop{}\!\mathrm{d}}

\makeatletter
\renewcommand{\section}[1]{\@startsection{section}{1}{\parindent}{8pt}{-5pt}{\normalfont\normalsize\bfseries}{#1.\ --}}
\makeatother

\begin{document}

\title{Packing of elastic wires in flexible shells}

\author{Roman Vetter}
\email{vetterro@ethz.ch}
\author{Falk K.~Wittel}
\author{Hans J.~Herrmann}

\affiliation{Computational Physics for Engineering Materials, IfB, ETH Z\"{u}rich, Stefano-Franscini-Platz 3, CH-8093 Z\"{u}rich, Switzerland}

\date{\today}

\begin{abstract}
The packing problem of long thin filaments that are injected into confined spaces is of fundamental interest for physicists and biologists alike. How linear threads pack and coil is well known only for the ideal case of rigid containers, though. Here, we force long elastic rods into flexible spatial confinement borne by an elastic shell to examine under which conditions recently acquired knowledge on wire packing in rigid spheres breaks down. We find that unlike in rigid cavities, friction plays a key role by giving rise to the emergence of two distinct packing patterns. At low friction, the wire densely coils into an ordered toroidal bundle with semi-ellipsoidal cross-section, while at high friction, it packs into a highly disordered, hierarchic structure. These two morphologies are shown to be separated by a continuous phase transition. Our findings demonstrate the dramatic impact of friction and confinement elasticity on filamentous packing and might drive future research on such systems in physics, biology and even medical technology toward including these mutually interacting effects.
\end{abstract}


\maketitle

\section{Introduction}

Dense filament packing can be found in various natural systems. A well known instance is the injection and subsequent coiling of long DNA in globules and viral capsids (e.g., \cite{RWC73,KTBG01,PH08}). A similar technique has been harnessed by neurosurgeons for the minimally invasive treatment of saccular aneurysms, into which detachable platinum wires are fed to initiate occlusion \cite{GVSM91}. Extremely dense fiber packing can also be observed for example in gland thread cells of hagfish \cite{F81}. Albeit their morphology has recently been unraveled \cite{WHMLDBBHVTF14}, the understanding of morphogenesis---the process of shape transformation and development---remains vague. In particular, the role of macroscopic material parameters is little understood.

Several experimental and numerical studies \cite{DGS02,DGS03,DOG06,DG07,GBCD08,SWH08} revealed how long wires form loop patterns and alignment between contacting segments when injected into rigid two-dimensional containers, depending on friction, plastic yield point and the precise insertion setup. In rigid three-dimensional cavities \cite{SW05,SNWHH11,NSWH12}, on the other hand, friction has been reported to have negligible impact on the packing process. In a recent study on morphogenesis of elastic ring filaments growing in flexible shells \cite{VWH14}, however, we discovered four morphological phases strongly dependent on friction, which brought it back into play in 3D. This naturally begs the question to which degree the knowledge previously acquired on the thread injection and packing problem in rigid cavities is applicable to less idealized, deformable ones, such as cell walls, vesicles, or arterial walls. In this paper, we perform a change of topology from ring-like (as in Ref.~\cite{VWH14}) to linear threads, which are more relevant in Nature and biomedical applications, to answer this question quantitatively using numerical simulations and simple table-top experiments. We show how confinement elasticity completely alters the packing process of a thin wire that is fed in. In strong contrast to rigid cavities, friction becomes the key macroscopic material property beyond a critical point, determining the packing process by giving rise to the spontaneous emergence of two different morphologies. We characterize them by comparing numerical measurements of their energetic and structural properties, finding that frictional forces lead to a highly disordered, hierarchic, crumpled structure governed by power laws, whereas absence of friction yields a dense toroidal bundle with semi-ellipsoidal cross-section. We further show that the transition from rigid to flexible cavities as well as the transition from weak to strong friction in flexible cavities is continuous and accompanied by spontaneous symmetry breaking, and we identify the associated order parameter. These results challenge researchers in the many fields where filamentous packing is relevant to carefully include the effects of friction and elasticity in their studies, and might trigger new efforts for instance in aneurysm coiling research, which has previously neglected these.

\section{Model}

For the theoretical considerations and numerical simulations, we model the thin wire by an extensible, intrinsically straight and untwisted Kirchhoff rod \cite{D92}. Its centerline, a space curve $\vec{x}(s)$, is parameterized by its arclength $s\in[0,L]$. In its deformed state, an orthonormal director frame $\{\vec{d}_i\}_{i=1,2,3}$ specifies the cross-sectional orientation along the curve, with the third one being the unit tangent $\vec{d}_3=\partial_s\vec{x}/\left\lVert\partial_s\vec{x}\right\rVert$. Its rate of change $\partial_s\vec{d}_i=\vec{k}\times\vec{d}_i$ defines the Darboux vector $\vec{k}$ with director components $k_i=\vec{k}\cdot\vec{d}_i$. Assuming a homogeneous, isotropic, linearly elastic material with Young's modulus $E_\mathrm{w}$ and Poisson's ratio $\nu_\mathrm{w}$, the elastic potential energy $U_\mathrm{w}$ of the wire reads
\begin{equation}
U_\mathrm{w} = \frac{1}{2} \int_0^L E_\mathrm{w}I\kappa^2 + GJk_3^2
+ E_\mathrm{w}A\epsilon^2\,\diff s,
\end{equation}
in which $\kappa=\sqrt{k_1^2+k_2^2}$ denotes the curvature, $k_3$ is the twist per unit length, whereas $\epsilon=\left\lVert\partial_s\vec{x}\right\rVert-1$ is the axial Cauchy strain due to compression or tension. For an invariant circular cross-section with radius $r$, $A=\pi r^2$ is the cross-section area, $I=\pi r^4/4$ the second moment of inertia, and $J=2I$ the polar moment of inertia. The shear modulus reads $G=E_\mathrm{w}/2(1+\nu_\mathrm{w})$.

Analogously, a Kirchhoff--Love shell with middle surface $\Omega$ and uniform thickness $t$ is used to represent the flexible hull. In linear elasticity (Young's modulus $E_\mathrm{s}$ and Poisson's ratio $\nu_\mathrm{s}$), the elastic potential energy of the shell reads \cite{K66}
\begin{align}
U_\mathrm{s} = \frac{1}{2} \int_{\Omega} H^{ijkl}\left( M \alpha_{ij}\alpha_{kl} + B \beta_{ij}\beta_{kl} \right)\,\diff\Omega,\\
H^{ijkl} = \nu_\mathrm{s}\overline{a}^{ij}\overline{a}^{kl}+\frac{1-\nu_\mathrm{s}}{2}(\overline{a}^{ik}\overline{a}^{jl}+\overline{a}^{il}\overline{a}^{jk}),
\end{align}
where $M=E_\mathrm{s}t/(1-\nu_\mathrm{s}^2)$ is the membrane rigidity and $B=Mt^2/12$ the bending rigidity. Einstein summation over repeated indices $i,j,k,l\in\{1,2\}$ is used, with subscripts (superscripts) denoting covariant (contravariant) coefficients. The membrane strains $\alpha_{ij}=(a_{ij}-\overline{a}_{ij})/2$ and the bending strains $\beta_{ij}=\overline{b}_{ij}-b_{ij}$ are the covariant coefficients of the change of first ($a$) and second ($b$) fundamental forms from the initial stress-free configuration (barred symbols) to the current one (bare symbols).

For a realistic treatment of body contacts between any combination of types (wire-wire, wire-shell, shell-shell), we use a contact model with dry stick-slip friction analogous to Refs.~\cite{SWH08,SNWHH11,VWH14} in the simulations. Repelling Hertzian contact forces are exchanged when two body volumes overlap, whereas the tangential contact forces are set to obey Coulomb's law with static and dynamic friction coefficients $\mu_\mathrm{s}$ and $\mu_\mathrm{d}$, respectively.

We minimize the elastic energies $U=U_\mathrm{w}+U_\mathrm{s}$ with the finite element method and integrate Newton's equations of motion with a common predictor-corrector scheme of second order with adaptive time-stepping. Subcritical viscous damping forces are added to allow the system to stay close to static equilibrium at all times. Details on these numerical models and their implementation can be found in Refs.~\cite{VWSH13,VSJWH13}.

An intrinsically straight thread is radially injected into a thin shell whose unstrained equilibrium configuration is a sphere with radius $R$ with a small opening with diameter $2r$ through which the thread is fed in. The confining shell is held in place by imposing zero displacement on a narrow rim about the entrance hole. A small random transverse deflection is initially added to the wire to break the rotational symmetry about the feeding axis. In all simulations, the following parameters are fixed: $r=0.5~\mathrm{mm}$, $E_\mathrm{w}=1~\mathrm{GPa}$, $\nu_\mathrm{w}=\nu_\mathrm{s}=1/3$. To reduce the influence of friction on the phase space to a single dimension, we always use a fixed ratio $\mu_\mathrm{d}/\mu_\mathrm{s}=0.9$. The wire is pushed in at a constant speed of $0.5~\mathrm{m/s}$, which is slow enough for inertial effects to have negligible impact on the outcome.

In summary, the system is effectively characterized by the four dimensionless and independent control parameters $\rho=R/r$, $\xi=R/t$, $\varepsilon=E_\mathrm{w}/E_\mathrm{s}$ and $\mu_\mathrm{s}$ if it is close to static equilibrium---a parameterization which is equivalent to that introduced in Ref.~\cite{VWH14}. The slenderness ratio $\xi$ and the cavity flexibility $\varepsilon$ generalize the rigid spheres considered in preceding related studies \cite{SW05,SNWHH11,NSWH12}.

\begin{figure}
	\includegraphics[width=\columnwidth]{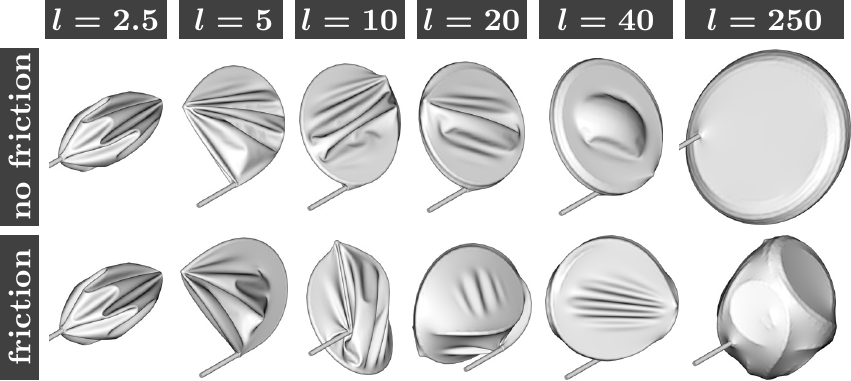}
	\caption{Packing evolution in flexible confinement. Series of simulation snapshots at $\mu_\mathrm{s}=0$ (top row) producing aligned coils, and at $\mu_\mathrm{s}=1/2$ (bottom row) producing disordered crumpled packings. The remaining parameters are $\rho=20$, $\xi=200$, $\varepsilon=100$. All images to scale.}
	\label{fig:flexible_packing}
\end{figure}

\section{Packing evolution and morphologies}

The typical packing evolution obtained in sufficiently elastic or thin shells is displayed in Fig.~\ref{fig:flexible_packing} for increasing thread lengths $l=L/R$, and a corresponding movie is provided in the supplementary material (see video\_s1.mpg). In absence of friction, a highly ordered packing pattern is observed in which the wire bundles into a tight toroidal coil (comparable, \textit{e.g.}, to how microtubules bundle up in erythrocytes \cite{WS91}, or to DNA spools in bacteriophage heads \cite{RWC73}) that continues to grow and stretch the confining shell as more thread is injected. Eventually, both upper and lower faces of the shell are fully flattened, turning it into the convex hull of the enclosed coil. Similar forms have been experimentally obtained by enclosing elastic nanotubes and nanowires with emulsion droplets and polymer shells, which were then intentionally contracted \cite{XWLYCWWLXC10,CWXSYZZZC11}. Monte Carlo simulations at finite temperature \cite{MO07,FIKM13} have likewise indicated that soft vesicles deform into obloids when enclosing a fluctuating polymer chain whose persistence length grows much larger than the vesicle diameter.

When friction is activated, however, the situation is dramatically changed: Tangential sliding is hampered, which lets the wire tip poke the surrounding shell significantly. The compressive forces acting on the inserted thread are much higher in consequence, and soon let it buckle out of the coiling plane to form a more three-dimensional packing process with frequent spontaneous loop reorientations, yielding a crumpled structure that tends toward a spherical globule at high packing density. Due to the thin shell's tensile flexibility, frictional forces let the pushing wire locally drag its confinement along a small distance, making static friction much more relevant in deformable than in rigid cavities.

\begin{figure}
	\includegraphics[width=\columnwidth]{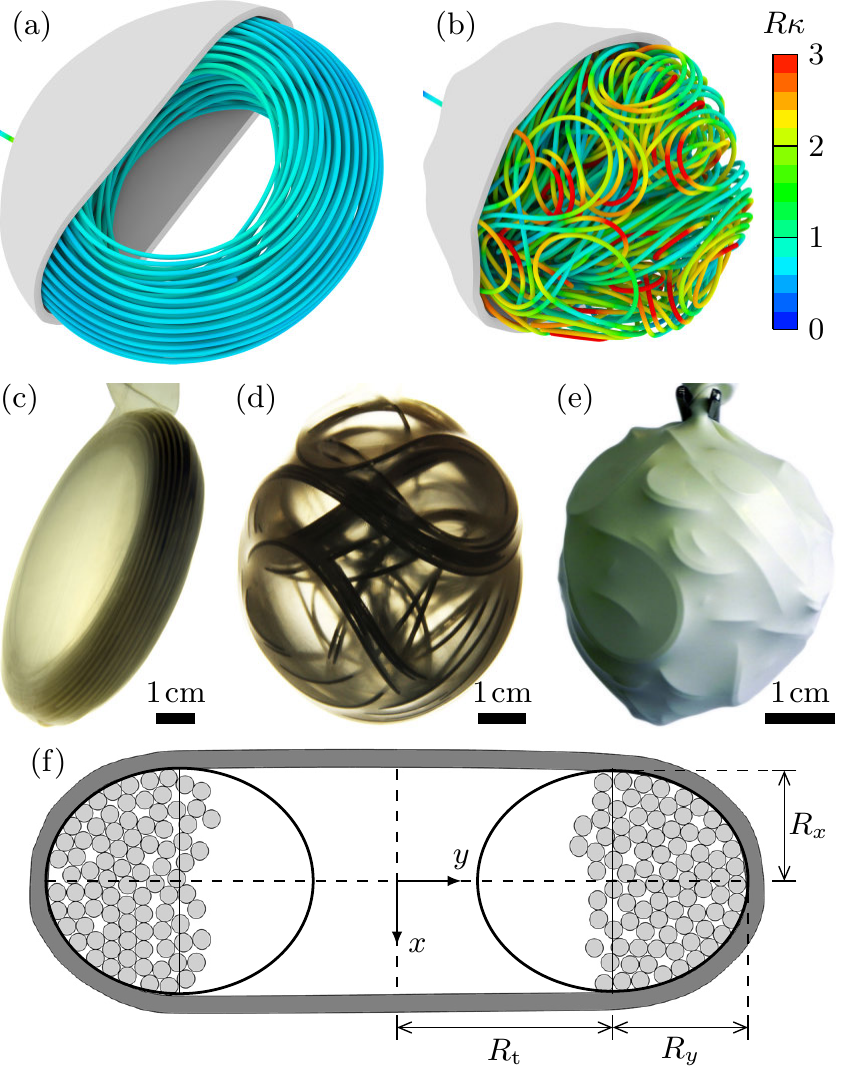}
	\caption{Packing morphologies in flexible confinement. (a) Toroidal coiling at weak friction ($\mu_\mathrm{s}=0$). (b) Crumpling at strong friction ($\mu_\mathrm{s}=1/2$). The further simulation parameters are $\rho=20$, $\xi=10$, $\varepsilon=10^3$, and the packed wire length is $l=800$. In color, the dimensionless curvature is shown. Images to scale, except that the wire radius is halved to reveal the inner structure. Movies for (a) and (b) are available in the supplementary material (video\_s2.mpg and video\_s3.mpg). (c--e) Experimental realizations of the two morphologies in transparent and opaque rubber balloons. (c) $r=1~\mathrm{mm}$, $R=23~\mathrm{mm}$, $l\approx950$. (d) $r=0.5~\mathrm{mm}$, $R=23~\mathrm{mm}$, $\mu_\mathrm{s}\approx0.55$, $l\approx650$. (e) $r=0.5~\mathrm{mm}$, $R=18~\mathrm{mm}$, $\mu_\mathrm{s}\approx1.2$, $l\approx 750$. (f) Cross-section of (a) with superposed ellipses and geometric parameters.}
	\label{fig:morphologies}
\end{figure}

For validation and demonstration, we have reproduced these two morphologies in table-top experiments, three examples of which are shown in Fig.~\ref{fig:morphologies}(c--e). Polycaprolactam wires ($r=0.5,1~\mathrm{mm}$, up to $L=22~\mathrm{m}$) were manually pushed through a straight steel pipe into customary inflatable balloons made of natural rubber ($R=18,23~\mathrm{mm}$, $t=0.25~\mathrm{mm}$). The friction coefficient was determined by placing pieces of wire on pieces of the rubber shell and measuring the tilt angle at which the wire started to slide. To reduce friction, we coated the wires with an acrylic dye and additionally used a silicone lubricant. These experiments yielded the same two morphologies as our numerical simulations.

\section{Analytical model for the coiled morphology}

To quantitatively characterize the coiled morphology at low friction, we measured the geometrical properties of cross-sections, of which a typical example with fully stretched lateral faces of the thin shell is displayed in Fig.~\ref{fig:morphologies}(a,f). Within the reasonable parameter range examined here ($20\leq\rho\leq40$, $10\leq\xi\leq200$, $1\leq\varepsilon\leq10^4$), if the shell is flexible enough, the cross-sections of the toroidal bundle are very well approximated by two half ellipses, with major toroidal radius $R_\mathrm{t}$, and two minor ellipsoidal radii $R_x$, $R_y$, as labeled in Fig.~\ref{fig:morphologies}(f). While the evolution of these three radii with the packed thread length depends on the system parameters $\rho$, $\xi$, $\varepsilon$, the common functional form is well approximated by a power-law scaling $R_\mathrm{t}/R\sim l^\alpha$, $R_x/R\sim l^\beta$, $R_y/R\sim l^\gamma$ in all observed cases within a certain range of the rescaled thread length $l=L/R$. This is demonstrated in Fig.~\ref{fig:scaling}(a) on an example simulation at $\rho=20$, $\xi=10$, $\varepsilon=10^3$ for which $\alpha=0.04\pm0.01$, $\beta=0.35\pm0.01$, $\gamma=0.46\pm0.01$, indicating that while the minor radii grow quickly, the major radius $R_\mathrm{t}$ remains approximately constant. $\alpha=0$ is indeed expected in the flexible cavity limit $M\sim E_\mathrm{s}t\to0$ (\textit{i.e.}, $\xi,\varepsilon\to\infty$), since $1/R_\mathrm{t}$ is the maximum curvature in the coil, the square of which is minimized according to the principle of minimum energy. This high degree of order allows for an analytical approximation following the ideas of Purohit \textit{et al.}~\cite{PKP03}. Their geometrical model, which was originally devised for the description of DNA coils in viral capsids, has also been successfully applied to ordered wire packing in rigid cavities \cite{SNWHH11,VWSH13}. It is based on the approximation that thin threads coil such that their binormal vectors are always parallel to the main coiling axis $x$, and that their radius of curvature about this axis is maximal, resulting in an empty cylindrical region as it is observed here. Assuming that the centerlines of individual strands in the coil are separated by a distance $d(L)\geq 2r$ on average, the number of windings along the $x$ axis is given by $w(y)=2R_x\sqrt{1-([y-R_\mathrm{t}]/R_y)^2}/d(L)$ for $y\geq R_\mathrm{t}$. The packed thread length follows as
\begin{equation}
L = \frac{2}{\sqrt{3}d(L)} \int_{R_\mathrm{t}}^{R_\mathrm{t}+R_y} \!\!2\pi y\,w(y)\,\diff y = \frac{2\pi cR_xR_y^2}{\sqrt{3}d(L)^2}
\label{eq:flexible_length}
\end{equation}
where $c=4/3+\pi p$ and $p=R_\mathrm{t}/R_y$. This fixes the average strand separation $d(L)$ given that $L,R_\mathrm{t},R_x,R_y$ are known. Analogously, the wire bending energy $U_\mathrm{b}=1/2\int_0^L E_\mathrm{w}I\kappa^2\,\diff s$ reads
\begin{equation}
U_\mathrm{b}(L) = \frac{2}{\sqrt{3}d(L)} \int_{R_\mathrm{t}}^{R_\mathrm{t}+R_y} \frac{\pi E_\mathrm{w}I}{y}\,w(y)\,\diff y.
\end{equation}
Using Eq.~(\ref{eq:flexible_length}), the unknown $d(L)$ can be eliminated to yield
\begin{equation}
U_\mathrm{b}(L) = \frac{2LE_\mathrm{w}I}{cR_y^2}\left(q\ln\left[\frac{1+q}{p}\right]+\frac{\pi}{2}p-1\right)
\label{eq:flexible_purohit_ubend}
\end{equation}
where $q=\sqrt{1-p^2}$. Note that Eq.~(\ref{eq:flexible_purohit_ubend}) is real-valued even for $p>1$. $U_\mathrm{b}$ is monotonically decreasing in $p$, and thus the wire favors toroidal configurations with large major radius $R_\mathrm{t}$ but small minor radii $R_x$, $R_y$. This tendency of the wire competes with the shell deformations necessary to adopt such a shape.

\section{Quantitative comparison of the two morphologies}

\begin{figure}
	\includegraphics[width=\columnwidth]{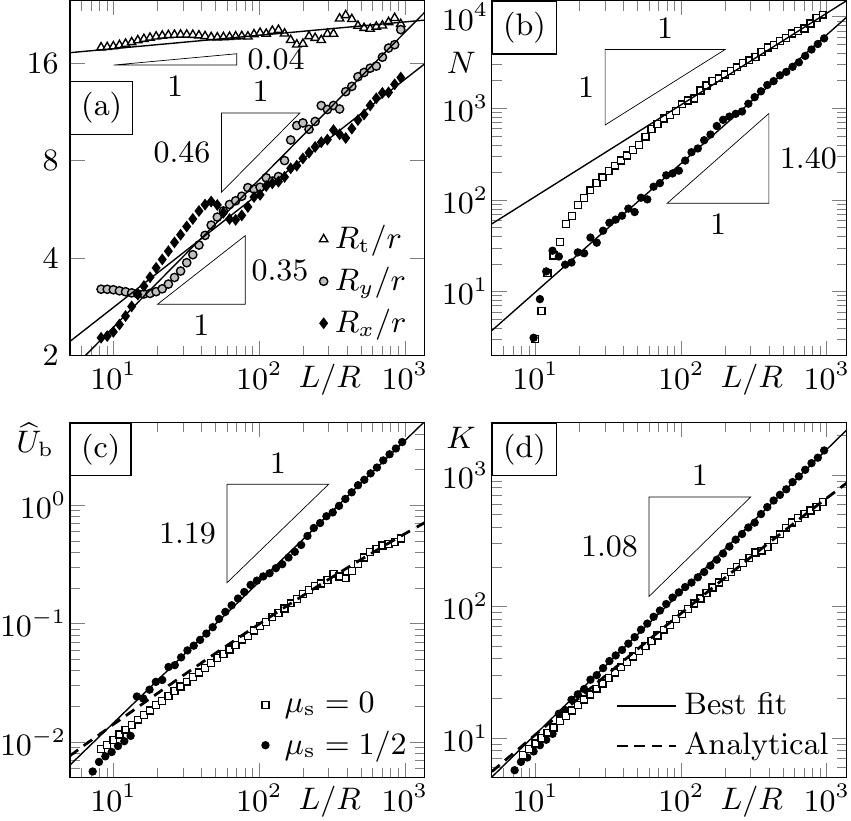}
	\caption{Quantitative comparison of the two morphologies. (a) Scaling of the shape of the coil with the inserted wire length. Data obtained by fitting semi-ellipses to the cross-sections as shown in Fig.~\ref{fig:morphologies}(f). (b) Number of discrete wire self-contacts $N$. (c) Non-dimensionalized bending energy $\widehat{U}_\mathrm{b}$. (d) Total curvature $K$. Symbol and line legends in subplots (c) and (d) hold for (b--d). Data from simulations at $\rho=20$, $\xi=10$, $\varepsilon=10^3$.}
	\label{fig:scaling}
\end{figure}

In Fig.~\ref{fig:scaling}(c), the dimensionless bending energies of both morphologies are compared. Since $U_\mathrm{b}\sim \rho^2E_\mathrm{w}I/R$ in the rigid shell limit \cite{SNWHH11}, we rescaled it to $\widehat{U}_\mathrm{b}:=U_\mathrm{b}r^2/RE_\mathrm{w}I$. The numerical measurement of the coiled morphology is very well approximated by Eq.~(\ref{eq:flexible_purohit_ubend}), with only a slight overestimation at early packing. The crumpled morphology simulated at $\mu_\mathrm{s}=1/2$, on the other hand, emerges by elastic bifurcation, which can be recognized in Fig.~\ref{fig:scaling}(c) by the early discontinuity in $\widehat{U}_\mathrm{b}$. It exhibits a clear power-law trend $\widehat{U}_\mathrm{b} \sim l^\delta$ with $\delta=1.192\pm0.006$, which hints at a hierarchic inner packing structure: The more wire is injected, the more spatial freedom is limited, resulting in higher bending curvature in the newly formed loops. The bulges visible in Fig.~\ref{fig:morphologies}(e) provide some intuition on this phenomenon. Such a power-law scaling of the bending energy discriminates the disordered phase in elastic shells reported here from that obtained in rigid spheres with intrinsically curved wires \cite{SNWHH11}.

To gain further insight into the packed structure, we consider the total curvature $K=\int_0^L\kappa\,\diff s$, which reads
\begin{equation}
K(L) = \frac{2}{\sqrt{3}d(L)} \int_{R_\mathrm{t}}^{R_\mathrm{t}+R_y} 2\pi\,\omega(y)\,\diff y = \frac{\pi L}{cR_y}
\end{equation}
in our coiling model (recall $c=c(L)$ as defined below Eq.~(\ref{eq:flexible_length})). This excellently approximates the numerical measurement with a slight initial overestimation, as shown in Fig.~\ref{fig:scaling}(d). In the hierarchic crumpled morphology at $\mu_\mathrm{s}=1/2$, we find a power law $K \sim l^\lambda$ with $\lambda=1.083\pm0.004$ consistently with the energetics. Again, a pronounced discontinuity in $K$ marks the beginning of its existence at the bifurcation point. The hierarchic nature of the crumpled structure is manifest also in the number of contacts between discrete wire segments (\textit{in silico}: finite elements), $N$, as Fig.~\ref{fig:scaling}(b) shows. After the two morphological phases have bifurcated, it follows a power law $N \sim l^\tau$ with $\tau=1.40\pm0.01$ at $\mu_\mathrm{s}=1/2$. On the contrary, low friction leads to dense alignment with hexagonal packing in the coiled morphology, hence $N\sim l$ once a few coil windings are established.

\section{Hierarchic disorder}

Above, we have termed the disordered morphology \textit{hierarchic}, based on the observed power-law scaling of the contact number, bending energy and total curvature. Now we consider the geometric structure of the packed wire in additional detail to put this statement on more quantitative ground.

\begin{figure}
	\includegraphics[width=\columnwidth]{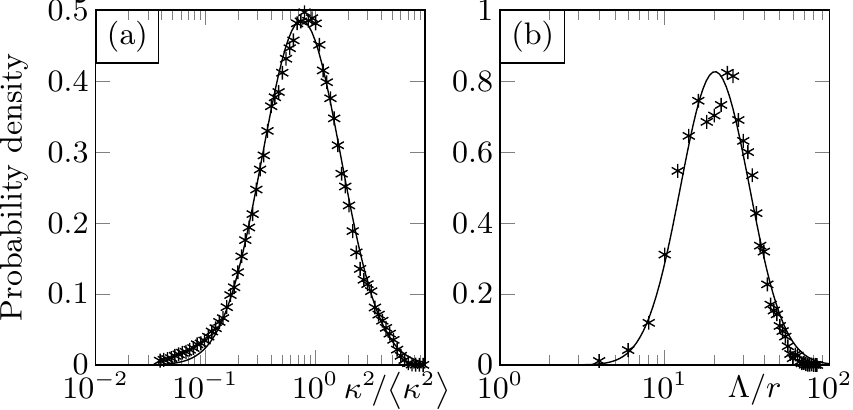}
	\caption{Hierarchy in the disordered morphology. (a) Distribution of the squared curvature $\kappa^2/\left\langle\kappa^2\right\rangle$. (b) Distribution of loop sizes $\Lambda/r$. Asterisks denote numerical data points, solid lines are best-fitting lognormal probability densities.}
	\label{fig:hierarchy}
\end{figure}

In rigid spheres, the distribution of bending energies in disordered wire packings is well fit by a lognormal probability density, with the exception of a larger tail toward low values \cite{SNWHH11}. The same holds here for the crumpled morphology in elastic shells, albeit with a different origin. In rigid spheres, such hierarchic disorder needed to be introduced by artificially pre-curving the wire, whereas tight membrane confinement induces this hierarchy naturally even with straight threads. Figure~\ref{fig:hierarchy}(a) shows that the logarithm of the normalized local squared curvature follows a normal distribution:
\begin{equation}
\ln\left(\frac{\kappa^2}{\left\langle\kappa^2\right\rangle}\right) \sim \mathcal{N}(\overline{x},\sigma^2)
\end{equation}
with probability density function
\begin{equation}
f(x) = \frac{1}{\sqrt{2\pi}\sigma}\exp\left[-\frac{1}{2}\left(\frac{x-\overline{x}}{\sigma}\right)^2\right].
\label{eq:curv_prob}
\end{equation}
Our numerical data for the crumpled morphology is best fit by Eq.~(\ref{eq:curv_prob}) with a mean value of $\overline{x}=-0.29$ and a standard deviation of $\sigma=0.83$.

Lognormal distributions are typical for hierarchic events. In crumpled thin sheets, for instance, the ridge lengths are lognormally distributed, see Ref.~\cite{W07} and the literature cited therein. For the present crumpled wires in elastic shells, the relevant local geometrical features are individual loops formed in the packing process. A wire loop is easily defined in two dimensions \cite{DGS02}, but in 3D it is somewhat less obvious. We define it here as the wire segment between two neighboring local maxima of the squared curvature $\kappa^2$ along the wire centerline, and denote its length by $\Lambda$. As shown in Fig.~\ref{fig:hierarchy}(b), the distribution of loop sizes in the crumpled morphology is lognormal in good approximation: $\ln\left(\Lambda/r\right) \sim \mathcal{N}(\overline{x},\sigma^2)$ with $\overline{x}=3.0$, $\sigma=0.48$.

\section{Order-disorder transition}

\begin{figure}
	\includegraphics[width=\columnwidth]{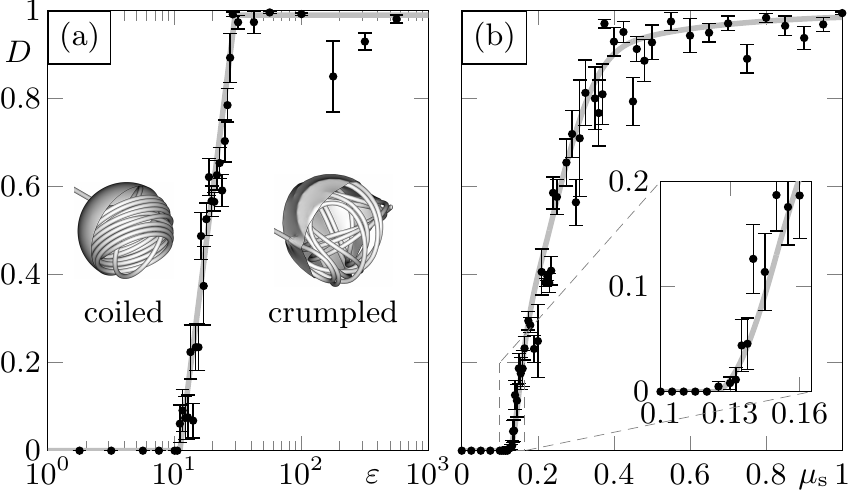}
	\caption{Order parameter $D$ for the coiling-to-crumpling transition. (a) As a function of the confinement flexibility $\varepsilon$ at $\rho=20$, $\xi=30$, $\mu_\mathrm{s}=1$. (b) As a function of the friction coefficient $\mu_\mathrm{s}$ at $\rho=20$, $\xi=10$, $\varepsilon=10^3$. The inset shows a magnification about the critical point. The solid gray lines are used as a guide to the eye. Error bars indicate standard errors from $10$ independent simulations. Data maximized over $l\in[5,50]$.}
	\label{fig:order_param}
\end{figure}

To complete the quantitative characterization of the two morphological phases, an order parameter is required that discriminates them rigorously. What defines the order of the coil at low friction is alignment between the individual loops. The hierarchic packing pattern in the crumpled morphology emerges due to frequent three-dimensional loop reorientations, which break this alignment. Denote by $\vec{n}$ the unit vector pointing in direction of the wire's principal axis of minimal moment of inertia. The curvature of the wire about $\vec{n}$ is given by the triple product $\kappa_{\vec{n}}(s)=\vec{n}\cdot(\partial_s\vec{x}\times\partial_s^2\vec{x})$, whose average sign $S=1/L\int_0^L\sgn\kappa_{\vec{n}}(s)\,\diff s$ measures the fraction of the wire turning in either direction about that axis. $S$ vanishes iff left- and right-turning wire segments are balanced (\textit{e.g.}, when the loops are isotropically distributed, degenerate $\vec{n}$), whereas it takes one of its extreme values $\pm1$ iff the coil never changes orientation. Since this coiling direction is initially selected at random, the sample average of $S$ is zero. The decisive non-trivial quantity is thus $\lvert S\rvert$, and $D=1-\lvert S\rvert$ is an order parameter that discriminates ordered coiling from disordered crumpling. A closely related quantity has already served to distinguish morphologies of growing elastic rings confined in spheres \cite{VWH14}. As shown in Fig.~\ref{fig:order_param}, $D$ vanishes in the coiled phase observed in relatively stiff confinements even with very strong friction, and in flexible shells if the friction coefficient is sufficiently small. Evidently, the crumpled phase ($D>0$) requires a flexible shell and moderate friction to emerge, and the phase transition is continuous.

\section{Conclusion and significance}

Our findings accentuate the fundamental difference between rigid and flexible cavities for the wire packing problem. Friction is responsible for a morphological phase transition from ordered coiling to a disordered hierarchic structure. In rigid spheres, such disorder needed to be introduced by artificially pre-curving the wire \cite{SNWHH11}, whereas in the tight flexible confinement considered here it occurs spontaneously even with straight rods. This insight might contribute to explaining the high degree of order observed in hagfish gland cell thread morphogenesis. It has far-reaching implications for biomedical applications such as the surgical occlusion of cerebral aneurysms, where the friction coefficient is estimated to be well above the critical value reported here and where common simulations ignore deformations of the arterial walls (\textit{e.g.}, \cite{BCGCF13}). While friction can cause problems during injection \cite{SF06}, our results suggest that it is nevertheless highly desirable in order to obtain a broader spatial distribution of loops, which is known to assist in stabilization \cite{BCGCF13}.

Elastic threads with circular topology develop four different morphologies when flexibly confined \cite{VWH14}. We have found in this letter that only two prevail when linear threads are injected instead. The reason is that elastic rings possess an instability when confined in spherical shells that linear threads do not possess: They are (initially) in full contact with the inner surface of the shell and can either buckle away from it or not. They remain on the two-dimensional surface in rigid spheres and are enabled to pack in a three-dimensional fashion in flexible shells. Linear wires, on the other hand, are never in full contact with the shell, due to their moment-free ends and the injection at a right angle. They hence avoid being trapped in the energetically unfavorable surface packing configuration, avoiding the instability in the first place.

An open question is the sensitivity of the packing behavior to material nonlinearity. It could also be worthwhile to determine the functional dependence of the exponents and critical points reported here on the system parameters to obtain a full image of the four-dimensional parameter space.

\begin{acknowledgments}
Financial support from the European Research Council (ERC) through Advanced Grant No.\ 319968-FlowCCS and from ETH Z\"{u}rich through ETHIIRA Grant No.\ ETH-03 10-3 is gratefully acknowledged.
\end{acknowledgments}

\end{document}